\documentclass{elsart}

\usepackage{graphics}

\usepackage{amssymb}

\begin{document}

\begin{frontmatter}


\thanks[label1]{This work is supported in the UK by EPSRC GR/R67699
(RealityGrid).
IP would like to acknowledge support from the European
Community Research Infrastructure Action under the FP6 Structuring
the European Research Area programme, as part of the HPC-Europa
project (contract number RII3-CT-2003-506079). Thanks to Paul Stansell
and Mike Cates for useful discussions.}

\title{Lattice Boltzmann for Binary Fluids with Suspended
Colloids}


\author[epcc]{K. Stratford}
\author[phys]{R. Adhikari}
\author[barca]{I. Pagonabarraga}
\author[epcc]{J.-C. Desplat}
\address[epcc]{Edinburgh Parallel Computing Centre and}
\address[phys]{School of Physics, James Clark Maxwell Building,
The University of Edinburgh,
The King's Buildings, Edinburgh EH9 3JZ, Scotland}
\address[barca]{Departament de F\'isica Fonamental, Universitat
de Barcelona, Carrer Mart\'i i Franqu\'es 1, 08028-Barcelona, Spain}

\begin{abstract}
A new description of the binary fluid problem via the lattice
Boltzmann method is presented which highlights the use of the
moments in constructing two equilibrium distribution functions.
This offers a number of benefits,
including better isotropy, and a more natural route to the inclusion of
multiple relaxation times for the binary fluid problem. In addition, the
implementation of solid colloidal particles suspended in the
binary mixture is addressed, which extends the solid-fluid
boundary conditions for mass and momentum to include a single
conserved compositional order parameter. A number of simple
benchmark problems involving a single particle at or near a
fluid-fluid  interface are undertaken and show good agreement
with available theoretical or numerical results.
\end{abstract}

\end{frontmatter}

\section{Introduction}
\label{}

The lattice Boltzmann equation (LBE) offers a very attractive
way to study complex fluid flows governed by the Navier-Stokes
equations
(for a recent review see, e.g.,
\cite{s02}). This is particularly the case for flows involving
irregular and/or changing geometry, e.g., flows in porous media,
multiphase systems, or multicomponent fluid mixtures. In the
case of fluid mixtures where flows can have highly convoluted
evolving structure, the LBE approach avoids the need to track
the time evolution of the interfaces between two or more components ---
pinch-offs and topological reorganisations occur naturally on
the lattice.
The situation becomes more complex still if solid particles,
or colloids, are suspended in the flow. Such colloidal suspensions
exhibit a range of interesting properties, and are of considerable
interest in many technological
and industrial applications (e.g., the food, cosmetic, and
pharmaceutical industries).

A number of LBE approaches to multiphase and multicomponent
fluids have been advanced \cite{swift,luo}, which adopt different
strategies to capture the important
thermodynamic interactions which give rise to surface tension
between the phases. While numerical LBE results for single
phase flow are very robust, multiphase flow is more problematic:
if the thermodynamics is not consistent, spurious unphysical
currents can be generated in interfacial regions owing to lack of
detailed balance.
While a complete, thermodynamically consistent, LBE approach to
the binary fluid problem is still an open issue,
the current methods can be used to provide useful results for
many problems of interest, provided the errors incurred are monitored
carefully, e.g., for phase separation dynamics \cite{vivprl,viv} and
droplet motion and break-up in shear flow \cite{droplet}. An alternative
approach \cite{xu} is to use the LBE in the momentum sector,
but employ a standard finite difference technique for the
evolution of a scalar order parameter describing the
composition.

Solid-fluid boundary conditions for stationary objects
of complex geometry are readily implemented within
the LBE. In addition, a general method
for the inclusion of moving solid objects in a single phase
is available
\cite{ladd94} which allows the representation of, e.g.,
moving spheres \cite{ladd96} and ellipsoids \cite{ellipsoid}.
These objects
are generally resolved on the fluid lattice (they are at least
several lattice spacings in size) without the need for
complicated curved boundary treatments.
Long range hydrodynamic interactions
between particles are captured by the LBE and, even when particles
are separated by distances small compared to the lattice size,
potentially important short-range lubrication interactions can
be included for simple objects such as spheres by 
adding back any unresolved contribution using an
analytical expression \cite{nguyen,ding}.

In this work, a number of these different strands are combined and
extended to address a new problem: that of colloids in a binary
solvent. First, a binary fluid LBE following \cite{swift} is
adopted in which the equilibrium distributions are cast in a
different, and more intuitive form. This leads to a number of
improvements in the behaviour of fluid-only problems. Second,
solid-fluid boundary conditions are extended to include colloidal
particles.  As one important advantage of the LB
approach is the relative ease with which solid-fluid boundary
conditions are included, it is useful to retain LB in both the
momentum and thermodynamic sectors. The boundary conditions for the
two sectors can then be implemented in a consistent fashion.

The paper is organised as follows. The following section gives an
overview of the basic LBE algorithm used and the extension of
solid-fluid boundary conditions for moving particles to the
order parameter sector. Section~3 presents a number of results
for simple test problems for the fluid only, and including a
single particle.

\section{The Lattice Boltzmann Approach}

In this section, the LB approach to the binary fluid problem
is described. As this differs from previous approaches \cite{swift,viv},
so the formalism is first set out for the lattice-Boltzmann
equation (LBE) for the single phase in a way
which makes clear the extension to the lattice kinetic equation
(LKE) describing the second phase.

\subsection{LBE for Navier-Stokes}

The LBE is commonly used to solve the conservation laws for
mass and momentum describing the hydrodynamics of an isothermal
fluid
\begin{equation}
\partial_t \rho + \nabla . \mathbf{g} = 0
\end{equation}
and
\begin{equation}
\partial_t \mathbf{g} +\nabla . \Pi = 0.
\end{equation}
These equations describe the dynamics of the mass density $\rho$
and the momentum density $\mathbf{g} = \rho \mathbf{v}$ for fluid
velocity $\mathbf{v}$. The momentum density can be written as the
mass flux $g_\alpha$ and the momentum flux as 
$\Pi_{\alpha\beta}$ so that for a Newtonian fluid
\begin{equation}
\Pi_{\alpha\beta} = g_\alpha v_\beta + p\delta_{\alpha\beta}
- \eta (\nabla_\alpha v_\beta + \nabla_\beta v_\alpha)
- \zeta \delta_{\alpha\beta} \nabla_\beta v_\beta,
\end{equation}
where Greek subscripts denote Cartesian components.
Together, these equations lead to the Navier-Stokes equations
for isothermal flow, namely
\begin{equation}
\rho (\partial_t \mathbf{v} + \mathbf{v}.\nabla\mathbf{v}) =
-\nabla p + \eta\nabla^2 \mathbf{v} + \zeta\nabla(\nabla . \mathbf{v}),
\end{equation}
where $p$ is the pressure, while $\eta$ is the shear viscosity and
$\zeta$ is the bulk viscosity.

The Lattice Boltzmann equation is derived from the (continuous time)
discrete velocity equation
\begin{equation}
\partial_t f_i + \mathbf{c}_i . \nabla f_i = - \mathcal{L}_{ij}(f_j - f^{eq}_j)
\label{eq:lbe}
\end{equation}
where the right-hand side corresponds to a linearised
collision operator. The discrete velocity set $\{\mathbf{c}_i\}$
are the nodes of a Gauss-Hermite quadrature \cite{luo,shanhe}
which ensure that the conserved moments of the distribution
function, $f_i$, have the same values as in the continuum. For the
isothermal LBE, these are the mass and momentum densities
\begin{eqnarray}
\rho &=& \sum_i f_i, \\
g_\alpha &=& \sum_i f_i c_{i\alpha}.
\end{eqnarray}
The Boltzmann kinetic description is restricted to a dilute gas
with an ideal equation of state $p = nk_BT = \rho c_s^2$, where
$n$ is the number density and $c_s$ is the isothermal speed of
sound. It is convenient to subtract the trivial kinetic contribution
to the pressure from the momentum flux tensor to define a deviatoric
momentum flux
\begin{equation}
S_{\alpha\beta} = \Pi_{\alpha\beta} - \rho c_s^2 \delta_{\alpha\beta}
= \sum_i f_i Q_{i\alpha\beta}
\end{equation}
where $\Pi_{\alpha\beta} = \sum_i f_i c_{i\alpha}c_{i\beta}$ is the
momentum flux
and $Q_{i\alpha\beta} = c_{i\alpha}c_{i\beta} - c_s^2\delta_{\alpha\beta}$
is referred to as the kinetic projector \cite{s02}.

In the most commonly used D$d$Q$n$ models, with $n$ velocities
(or quadrature nodes) in $d$ dimensions, the equilibrium
distribution functions are given by
\begin{equation}
f_i^{eq} = w_i \Bigg[ \rho + \frac{\rho v_\alpha c_{i\alpha}}{c_s^2}
+ \frac{\rho v_\alpha v_\beta Q_{i\alpha\beta}}{2c_s^4} \Bigg]
\label{eq:fequilibrium}
\end{equation}
where the $w_i$ are weights defining the quadrature and repeated
Greek indices are understood to be summed over. This form is obtained
by a truncation of the Hermite polynomial expansion of the 
Maxwell-Boltzmann distribution at second order \cite{shanhe}.
Finally, the BGK
collision matrix $\mathcal{L}_{ij}$ must satisfy the constraints
imposed by the conservation laws and rotational symmetry. In the
single relaxation time approximation $\mathcal{L}_{ij} = \delta_{ij} / \tau$,
and the shear and bulk viscosities are related to the relaxation time
$\tau$ by $\eta = \rho c_s^2 \tau$ and $\zeta = (2/d)\rho c_s^2 \tau$,
respectively. For multiple relaxation time models, the above relations
remain valid if $\tau$ is replaced by separate relaxation time for the
shear and bulk viscous stress.

The final LBGK equation is obtained by a further discretisation of
the continuous time equation. A second-order characteristic-based
method \cite{dellar} can be followed to obtain
\begin{equation}
f_i(\mathbf{r} + \mathbf{c}_i \Delta t; t + \Delta t)
- f_i (\mathbf{r}; t) = - (f_i - f_i^{eq}) /
(\tau + {\textstyle\frac{\Delta t}{2}}).
\label{eq:newlbe}
\end{equation}
This does not introduce lattice artefacts at second order so
that the definition of the viscosity remains as in the
continuum, unlike that in a first order scheme, where
the lattice error is absorbed into the viscosity to give
$\eta = \rho c_s^2 (\tau - 1/2)$.

\subsection{LKE for Cahn-Hilliard}

The starting point for the binary fluid mixture in this work is the
(Landau-Ginzburg-Wilson) free energy functional which describes the
total energy of a system
of fixed volume as a functional $F[\phi]$ of a single compositional order
parameter $\phi(\mathbf{r}, t)$. The order parameter measures the
ratio of the number density of particles of the two components of
the mixture $n_1$ and $n_2$ so that
$\phi = (n_1 - n_2) / (n_1 + n_2)$. The choice
of the free energy uniquely defines the physical quantities of
interest such as the fluid-fluid interfacial tension $\sigma$ and
the interfacial width $\xi_0$.
The chemical potential is obtained from the free energy functional
via
\begin{equation}
\mu = \frac{\delta F[\phi]}{\delta \phi}
\end{equation}
so that the thermodynamic force density acting on the fluid is
then $-\phi\nabla\mu$.

The equation of motion for the order parameter is the Cahn-Hilliard
equation
\begin{equation}
\partial_t \phi + \nabla . (\phi \mathbf{v} - M\nabla\mu) = 0.
\end{equation}
This is a conservation law involving the single conserved
quantity $\phi$, the flux of which $\phi\mathbf{v} - M\nabla\mu$
is made up of an advective component related to the fluid
velocity $\mathbf{v}$ and a diffusive component related to
the gradient of the chemical potential by an order parameter
mobility $M$. The mobility is assumed to be constant and
independent of $\phi$. A kinetic relaxation scheme for the
Cahn-Hilliard equation is obtained by introducing distributions
$g_i$ obeying the kinetic relaxation equation
\begin{equation}
\partial_t g_i + \mathbf{c}_i . \nabla g_i = - {\mathcal L}_{ij}^\phi
(g_j - g_j^{eq}).
\end{equation}
The discrete velocities ${\mathbf{c}_i}$ are the same as for
the LBE, and physical quantities are again related to moments
of the distribution via $\phi = \sum_i g_i$ and
$j_\alpha = \sum_i g_i c_{i\alpha}$. In choosing the form of
the equilibrium distribution, constraints similar to those
of \cite{swift} are employed, namely
\begin{eqnarray}
\sum_i g_i^{eq} = \phi, \label{eq:gm0} \\
\sum_i g_i^{eq} c_{i\alpha} = \phi v_\alpha, \label{eq:gm1}\\
\sum_i g_i^{eq} c_{i\alpha} c_{i\beta} = \mu\delta_{\alpha\beta}
+ \phi v_\alpha v_\beta. \label{eq:gm2}.
\end{eqnarray}
While the physical interpretation of the constraints on the
zeroth and first moments is clear, that for the second moment
is somewhat less so. There are two contributions, the first
of which ensures a diffusive contribution to the evolution of
the order parameter related to the chemical potential, while
the second represents an advective flux related to the velocity
field. It is then possible to write down an equilibrium
distribution by analogy with Eq.~\ref{eq:fequilibrium}, i.e.,
\begin{equation}
g_i^{eq} = w_i \Bigg[ \phi + \frac{j_\alpha c_{i\alpha}}{c_s^2}
+ \frac{(\mu\delta_{\alpha\beta} + \phi v_\alpha v_\beta
-\phi c_s^2 \delta_{\alpha\beta})
Q_{i\alpha\beta}}{2c_s^4} \Bigg].
\label{eq:gequilibrium}
\end{equation}
While this choice satisfies the above constraints, it is not
unique.  As it turns out, a slightly modified version is required
in practice (see following section).

Further discretisation again provides an LBGK equation with a
single relaxation time for the order parameter
\begin{equation}
g_i (\mathbf{r} + \mathbf{c}_i \Delta t; t + \Delta t)
- g_i (\mathbf{r}; t) = - (g_i - g_i^{eq}) /
(\tau^\phi + {\textstyle\frac{\Delta t}{2}}).
\label{eq:newlke}
\end{equation}
The order parameter mobility is related to the relaxation time
via $M = c_s^2 \tau^\phi$, which again can be adjusted to absorb
the discrete lattice correction. Note that the relaxation time
is not fixed in this approach, in contrast to previous work.

\subsection{Implementation}

The choice of free energy functional follows that of
\cite{viv}, where
\begin{equation}
F[\phi] = \int \mathrm{d}\mathbf{r} \big[
{\textstyle\frac{1}{2}}A\phi^2 +
{\textstyle\frac{1}{4}}B\phi^4 +
{\textstyle\frac{1}{2}}\kappa (\nabla\phi)^2 \big].
\end{equation}
This describes a symmetric binary mixture in which the
bulk free energy is related to two parameters $A$ and $B$,
while a term in a third parameter $\kappa$ penalises gradients
in the order parameter, i.e., it tends to minimise curvature of
the interface. The fluid-fluid interfacial tension is then
\begin{equation}
\sigma = (-8\kappa A^3 / 9 B^2)^{1/2}.
\end{equation}
If the further constraint that that $A=-B< 0$ is added, then
the order parameter lies predominantly on the interval $[-1,1]$. The
equilibrium interfacial profile is a \textit{tanh} with characteristic
width
\begin{equation}
\xi^{eq} = (2\kappa / A)^{1/2}.
\end{equation}

For the results presented in this work, the thermodynamic force
arising from the order parameter sector is introduced via a
correction to the equilibrium stress from a chemical pressure
tensor
\begin{equation}
P_{\alpha\beta} = 
\left[ \textstyle{\frac{1}{2}} A\phi^2 + \textstyle{\frac{3}{4}} B \phi^4
- \kappa\phi\nabla^2 \phi - \textstyle{\frac{1}{2}} \kappa (\nabla\phi)^2
\right] \delta_{\alpha\beta}
+ \kappa \nabla_\alpha \phi \nabla_\beta \phi.
\label{eq:pchem}
\end{equation}

Having made these choices, the lattice Boltzmann algorithm of collision
followed by propagation can be implemented in the normal way.
The post-collision distributions, denoted $f_i^\star$,
are based on Eq.~\ref{eq:fequilibrium} so that
\begin{equation}
f_i^\star = w_i \Bigg[ \rho^\star + \frac{g_\alpha^\star c_{i\alpha}}{c_s^2}
+ \frac{S_{\alpha\beta}^\star Q_{i\alpha\beta}}{2c_s^4} \Bigg]
\label{eq:fproject}
\end{equation}
where the density and momentum are unchanged by the collision
process so that $\rho^\star = \rho$
and $g_\alpha^\star = g_\alpha$. The deviatoric stress is relaxed
with a single relaxation time satisfying
$\eta = \rho c_s^2 \tau$
\begin{equation}
S_{\alpha\beta}^\star = S_{\alpha\beta} -
(S_{\alpha\beta} - S_{\alpha\beta}^{eq}) / (\tau +
{\textstyle\frac{\Delta t}{2}})
\end{equation}
where $S_{\alpha\beta}^{eq} = \rho u_\alpha u_\beta + P_{\alpha\beta}$.
Note that this reprojection of the physical properties to the distribution
readily admits the use of multiple relaxation
times, e.g., if separate values for the shear and bulk viscosities
are required.

For the order parameter, there is only one conserved quantity in the
collision $\phi^\star = \phi$, where again the star refers to post-collision
quantities. The order parameter flux is relaxed with single relaxation
time satisfying $M = c_s^2 \tau^\phi$ so that
\begin{equation}
j_\alpha^\star = j_\alpha - (j_\alpha - j_\alpha^{eq}) /
(\tau^\phi + {\textstyle\frac{\Delta t}{2}})
\end{equation}
with $j_\alpha^{eq} = \phi v_\alpha$. Finally, the values of $\mu$ and
$\mathbf{v}$ are known, allowing a reprojection of the
physical quantities to the $g_i^\star$ distributions.
If the equilibrium distributions Eq.~\ref{eq:gequilibrium}
are used the model is, in general, numerically unstable.
A simple finite-difference expansion \cite{fd} of the evolution equation
(not shown) shows a spurious up-gradient diffusion arises which
is the cause of the instability. Instead, it is possible to use
\begin{equation}
g_i^\star = \phi^\star \delta_{i0} + w_i
\Bigg[ \frac{j_\alpha^\star c_{i\alpha}}{c_s^2}
+ \frac{(\mu\delta_{\alpha\beta} + \phi^\star v_\alpha v_\beta)
Q_{i\alpha\beta}}{2c_s^2} \Bigg]
\end{equation}
where the $\delta_{i0}$ has the effect of moving most of the
order parameter into the non-propagating rest distribution $g_0$.
This choice is found to have very good stability properties,
while also satisfying the constraints on the moments of $g_i$.

\subsection{Colloidal particles}

A very general method for the representation of solid particles
within the LB approach was first put forward by Ladd \cite{ladd94}.
Solid particles (of any shape) can be defined by a boundary
surface which intersects a set of vectors $\{ \mathbf{c}_b \Delta t\}$
which connect lattice nodes inside and outside the surface. These
are referred to as boundary links. In the original approach, this
set of links defined a (spherical) shell with fluid occupying
all lattice nodes both inside and outside the ``solid'' object.
While this so-called internal fluid can be criticised as unphysical,
it actually exerts significant effect on the dynamics of the particle
in a single phase flow only on the time scale that it takes sound
waves to cross the particle \cite{heemels}. However,
in a binary fluid mixture,
it becomes essential to have truly solid particles to prevent
possible unphysical transfer of fluid across particle
surfaces. Such unphysical transfer would impact on the net
composition of the real fluid (i.e., that outside the particles).
This is particularly true for particles which wet one species
of fluid preferentially (non-neutral wetting), where a particle
might capture one species of fluid and hence unbalance the net
composition and/or generate spurious thermodynamic forces in
the region of the surface. A number of implementations of truly
solid particles have been developed \cite{aidun,heemels}; this work
extends that of Nguyen and Ladd \cite{nguyen} to the
binary fluid problem.

\begin{figure}
\begin{center}
\includegraphics{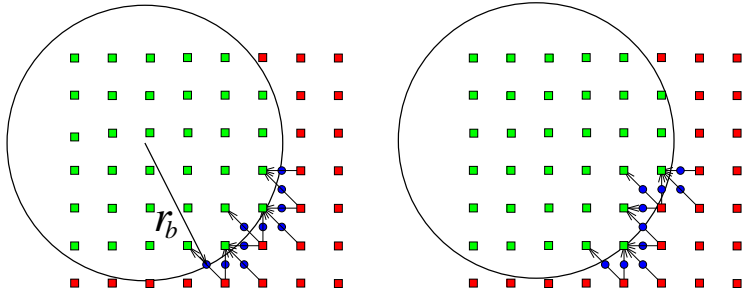}
\end{center}
\caption{A schematic showing the distribution of boundary links
for a section of a spherical particle. The nominal radius of the
particle is $a$, while the vector joining the centre to the
boundary nodes is $\mathbf{r}_b$. If the particle moves relative
to the lattice, fluid nodes may be exposed (right), in which
case fluid with appropriate properties must be added back. The
actual lattice used in calculations is D3Q15.}
\label{fig:schematic}
\end{figure}

A schematic diagram showing the distribution of links
for a section of a spherical
particle is shown in Figure~\ref{fig:schematic}. The centre of a sphere
of radius $a$ at $\mathbf{r}_0$ defines the position of the links,
and the particle moves smoothly across the lattice with linear
velocity $\mathbf{U}$
and angular velocity $\mathbf{\Omega}$. Boundary nodes are defined
to be half way along the links, which the set of vectors joining
the centre of the sphere to the boundary nodes denoted
$\{ \mathbf{r}_b \}$. For a single integral particle (having a
full complement of links)
\begin{equation}
\sum_b w_{c_b} \mathbf{c}_b = 0
\label{eq:wcb}
\end{equation}
and
\begin{equation}
\sum_b w_{c_b} (\mathbf{r}_b \times \mathbf{c}_b) = 0
\end{equation}
where $w_{c_b}$ are again the quadrature weights appropriate for the
boundary links. These results will be useful in the next section.

\subsubsection{Boundary conditions and particle dynamics}

The boundary condition developed by Ladd is usually referred to
as bounce-back on links (BBL). For a moving particle, this
takes the incoming distribution at the solid-fluid boundary,
$f_b$, and reflects it back along the incoming direction
\begin{equation}
f_{b'} = f_{b} - 2w_{c_b}\rho \mathbf{u}_b . \mathbf{c}_b / c_s^2
\label{eq:bbl}
\end{equation}
where $\mathbf{c}_{b^\prime} = -\mathbf{c}_b$. The correction is
related to the local solid body velocity at the boundary node
\begin{equation}
\mathbf{u}_b = \mathbf{U} + \mathbf{\Omega}\times\mathbf{r}_b,
\label{eq:ub}
\end{equation}
and has the effect of transferring mass from the leading edge of
the particle to the trailing edge. If the approximation is made
that $\rho \approx \rho_0$, i.e., that the density is approximately
constant around the particle, then Equations~\ref{eq:wcb}--\ref{eq:ub}
can be combined to show that the BBL applied around the whole
particle conserves mass.

Conservation of linear and angular momentum entails computing
the momentum transfer from the fluid to the particle as a
result of BBL, and updating the particle linear and angular
velocity appropriately. For this, the fully implicit method
described by Nguyen and Ladd \cite{nguyen} is used.

The BBL condition for density is extended to the order
parameter \cite{desplat} so that, at the surface,
the distributions $g_b$ are again reflected with a correction
\begin{equation}
g_{b'} = g_{b} - 2w_{c_b}\phi \mathbf{u}_b . \mathbf{c}_b / c_s^2.
\end{equation}
However, the order parameter $\phi$ around the particle
cannot, in general, be approximated by a constant as is the
case for density. This means that the order parameter is not
conserved at BBL by an amount
\begin{equation}
\delta\phi = 2\sum_b w_{c_b} \phi \mathbf{u}_b . \mathbf{c}_b / c_s^2
\end{equation}
for a given particle. To ensure the composition of the fluid
does not drift over time, the deficit or excess $\delta\phi$
is added back as a correction at the following time step:
\begin{equation}
g_{b'} = g_{b} - 2w_{c_b}\phi \mathbf{u}_b . \mathbf{c}_b / c_s^2
- w_{c_b}\delta\phi / \sum_b w_{c_b}.
\end{equation}
The size of the additional term is generally small compared
with the term in $\mathbf{u}_b . \mathbf{c}_b$, so this
represents a negligible perturbation to the motion.

Perhaps more serious are the corrections that arise when the
particle changes shape as it moves across the lattice. When
the particle motion exposes a fluid lattice node or recovers
a solid node, fluid with the appropriate properties must be
added or removed. For density, this gives
rise to a correction to the BBL Eq.~\ref{eq:bbl}
that ensures the mean fluid density $\rho_0$ does not
drift \cite{nguyen}. A similar correction can be made in the
order parameter sector which ensures that the mean fluid
composition $\phi_0$ does not change. For example, if fluid is added
at a newly exposed lattice node, then new properties are
determined by a linear interpolation of the distribution
functions at adjacent fluid sites. For order parameter,
an extra correction to the BBL at the following step
 of $\delta\phi = (\phi - \phi_0)$ is required
to maintain mean fluid composition, $\phi$ being the order parameter
added or removed. These corrections lead
to small unphysical fluctuations in the order parameter
near a moving colloid surface. At present, these are accepted
as the cost of preventing drift in the fluid composition.

Finally, the gradient in the order parameter field is required
to compute $P_{\alpha\beta}$ from Eq.~\ref{eq:pchem}. Near the
solid particles, these
gradients are computed following \cite{desplat} where the
order parameter is extrapolated along links to the boundary
nodes. In this work only neutral wetting is considered where
the contact angle between the solid and the fluid-fluid interface is 90
degrees. It should be noted that in imparting the thermodynamic
force to the fluid via the stress $P_{\alpha\beta}$, there is
no direct thermodynamic force on the colloid. The order parameter
only then affects the particle motion indirectly via the fluid
velocity fluid.

\subsubsection{Particles close to contact}

The BBL for the density distribution functions $f_i$ allows the
net hydrodynamic force and torque on a particle to be computed,
from which the particle velocity and position can be updated in turn.
For an isolated particle this is straightforward. However,
if two or more moving particles are close enough on the lattice
scale that there are no fluid sites in the interstice, the full
lubrication force between the particles will not be resolved.
In this case, it is possible to add back the unresolved part from
the analytical expressions available for pairwise forces between
spheres \cite{jeffrey} and, after appropriate calibration, recover
the correct lubricating behaviour at close approach \cite{nguyen}.

The same effect can occur in the order parameter sector when,
for example, two particles are close together at a fluid-fluid
interface. The particles should experience capillary forces
owing to the curvature of the interface in the gap between them,
but, again, if no fluid is present this force will not be
resolved. Unfortunately, there is no simple way to add back
the unresolved force as is done for the hydrodynamic
lubrication as the capillary force depends in a complex way
on the curvature of the interface in the vicinity of the
particles. While these capillary forces may be important for
some problems, they do not have the same impact as the
lubrication forces in, for example, preventing the particles
overlapping. However, the results presenting in the following
section only consider a single particle. Calibration of the
hydrodynamic radius of different particles for the single fluid
is carried out following \cite{nguyen}; these are assumed to
remain unchanged for the binary fluid.

\section{Results}

\subsection{Fluid only}

\begin{figure}
\begin{center}
\includegraphics{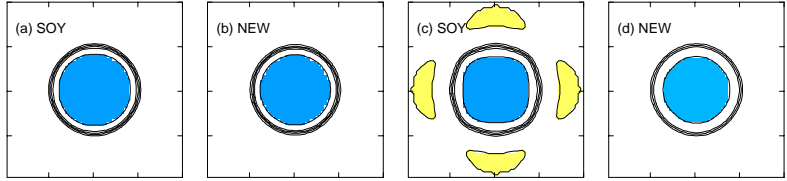}
\end{center}
\caption{An initially steady spherical interface between two fluids
is allowed to relax for some 100s of time steps. The LBE method described
in \cite{swift} (``SOY'') is compared with the new method. While the two
methods give similar results at surface tension 0.0042(a and b),
anisotropies are exposed using the SOY distributions at higher surface
tension 0.055 (c and d).}
\label{fig:drop}
\end{figure}

As a demonstration that the equilibrium distributions presented
in the previous section lead to better behaviour in the fluid
sector than those used previously, Fig.~\ref{fig:drop} shows a
comparison of results for a simple test problem. An initially
steady spherical fluid droplet
of one phase is initialised surrounded by the second phase. The
system, here $64^3$ lattice sites with periodic boundary conditions,
is then allowed to relax for a few hundred time steps. This problem
is repeated for two parameter sets taken from Table~3 of \cite{viv}
corresponding to surface tensions of $\sigma = 0.0042$ and
$\sigma = 0.055$. While there is little visible difference
between the results for the lower surface tension, anisotropy
in the old distributions is exposed at higher surface tensions
which do not infect the new distributions. This improved behaviour
is also manifest in improved numerical stability. While by no means
unconditionally stable, use of the new distributions allows a wider
range of parameter space to be investigated safely.

\subsection{Single particle at fluid-fluid interface}

A spherical particle subject to gravity may be suspended
at a fluid-fluid interface by interfacial tension. For a
particle which has a density difference $\Delta\rho$ with
the surrounding fluid, the key dimensionless number in this
situation is the Bond number
\begin{equation}
Bo = a^2 \Delta\rho g / \sigma
\end{equation} 
where $g$ is the gravitational acceleration. The Bond number
expresses the balance between the downward (or upward for
buoyant particles) force and the opposing interfacial
tension. If  $Bo << O(1)$ the interfacial tension can support
the particle with little deformation. As $Bo$ increases,
the interface bows downward until the particle can no longer
be supported. For a neutrally wetting particle with a contact
angle of 90$^o$ between solid and fluid-fluid interface,
the critical Bond number is 3/4; particles break away from
the interface for higher values.

In the LB model, a single particle of radius $a$ is placed
at rest across an initially flat interface and subject to a
downward force. The particle is allowed to fall until
it comes to rest, in which case the equilibrium displacement
below the level interface is measured,  or it detaches from
the interface. The width of the system in the horizontal
direction is at least $10a$ so that there is no significant
curvature of the interface at the periodic boundary conditions.
In addition, the second interface required in the system is
placed at sufficient distance that it does not affect the
motion of the particle.

\begin{figure}
\begin{center}
\includegraphics{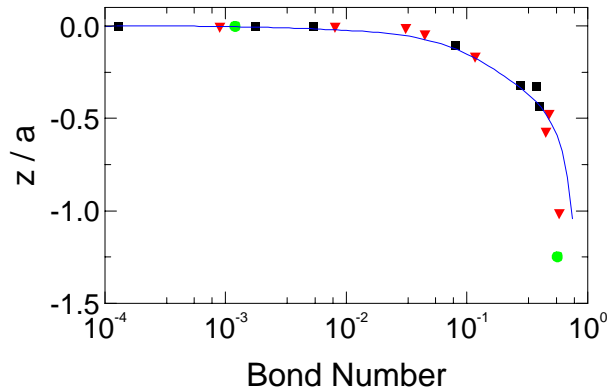}
\end{center}
\caption{The normalised equilibrium displacement $z/a$ of a single
particle below an initially flat interface as a function of Bond
number. The different symbols represent particles of different
radius: $a = 2.3$ circles, $a = 3.71$ triangles, and $a = 4.77$
squares. The curve is the theoretical result \cite{derjaguin46}
is valid for $Bo << 1$. The particle detaches from the interface
for Bond numbers higher than the critical value $B = 3/4$.}
\label{fig:bond}
\end{figure}

Figure~\ref{fig:bond} shows the normalised displacement $h/a$
as a function of Bond number for a number of different particle
sizes. For $Bo < 0.01$ the interface is essentially
rigid with no significant displacement, while
the displacement increases to a significant fraction of
the particle size for higher Bond numbers. The agreement
between the numerical results and the theoretical result
is generally excellent. As $Bo$ approaches the critical value
of $3/4$ the smaller particles drop somewhat below the
theoretical curve and can detach from the interface at
Bond numbers as low as $0.55$. This is likely to be caused
by a poor representation of the contact line for smaller
discrete particles.

\subsection{Particle approaching a fluid-fluid interface}

\begin{figure}
\begin{center}
\includegraphics{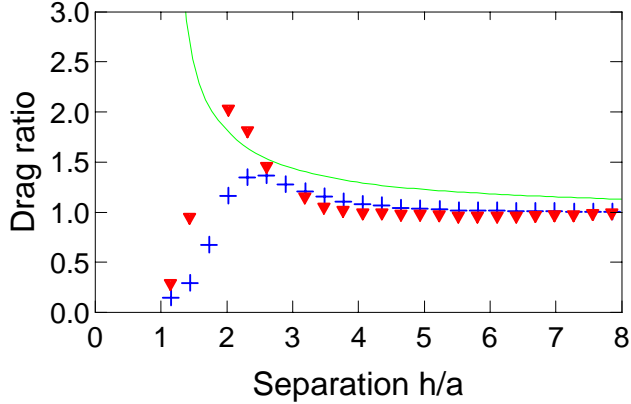}
\end{center}
\caption{The drag (normalised by $6\pi\eta a$) on a single sphere
sedimenting toward a stationary fluid-fluid interface as a function
of the normalised separation $h/a$. The solid line is based on
tabulated data from \cite{lee80}, while the symbols represent the
current model: triangles represent interfacial width
$\xi^{eq} = 0.8$ and crosses $\xi = 1.6$ lattice units.}
\label{fig:drag}
\end{figure}

As a second test of particle motion in a binary mixture,
the drag on a single sphere sedimenting toward a stationary
fluid-fluid interface is computed (Fig.~\ref{fig:drag}).
First, the mean drag on the sphere is obtained in a single
phase LB calculation to provide the calibration $6\pi\eta a$
as for the hydrodynamic radius (the nominal radius used here is $a = 2.3$).
The drag is then recomputed for
the same particle in the binary fluid as it sediments vertically
toward a horizontal fluid-fluid interface. The Bond number is
kept small enough that there is little deformation in the
interface as the particle approaches. The drag on the sphere
is measured as a function of the normal separation $h/a$; as
the exact value depends upon the position of the particle
relative to the lattice, the symbols in Fig.~\ref{fig:drag}
represent an average over 20 different particle configurations
and binned at intervals in $h/a$.

The results show that, far from the interface, the drag is
very similar to that seen in the single phase. This suggests
the redistribution of order parameter as the particle moves
does not strongly influence the dynamics. As the particle
approaches the interface, the drag increases but then decreases
sharply as the leading edge of the particle touches the
interface. The effect of differing interfacial width is
shown: the sharper interface induces a steeper rise in the
drag as the particle approaches. However, in both cases the
particle is finally captured by the interface.
For comparison, the exact results for a similar problem in
which a sphere approaches a flat interface of zero thickness
is shown \cite{lee80}. In this case, the particle never
actually reaches the interface.

\section{Closing Remarks}

This work has demonstrated the use of a new binary fluid lattice
Boltzmann approach applied to the problem of colloidal particles
The representation of colloidal particles within an LBE binary
fluid allows a large number of interesting physical problems to
be investigated. The results presented shown an excellent
agreement with exact results for a number of simple test
problems, and provides confidence that the approximations
made in the approach do not result in undue errors.

A number of potential improvements can be identified. First,
the thermodynamic force follows \cite{swift} in entering
via the equilibrium stress. This can be added directly as
an additional force on the fluid in the LBE. The implementation
of this additional force in the presence of solid particles
requires some care and is addressed elsewhere \cite{usinprep}.
Second, the problem of resolving near-contact capillary forces
would be most elegantly addressed via the use of a finite-volume
like approach where there is always some fluid retained in the
interstice between particles. This approach would also have the
benefit of preventing abrupt changes in the discrete shape of
a moving particle. However, this would represent a
considerable increase in complexity over the current approach.

In the meantime, the work demonstrates the flexibility of the
lattice Boltzmann method in addressing problems that are very
difficult to address at all, and currently intractable using
other methods.



\end{document}